\documentclass[twocolumn,floatfix,prb,eqsecnum,showpacs]{revtex4}
\usepackage{graphicx}
\usepackage{amsmath}
\usepackage{bm}
\begin{document}

\title{The gapped state of a carbon mono-layer in periodic magnetic and electric fields}

\author{I. Snyman}
\email{isnyman@sun.ac.za}
\affiliation{National Institute for Theoretical Physics, Private Bag X1, 7602 Matieland, South Africa,}
\affiliation{Department of Physics, Stellenbosch University, Private Bag X1, 7602 Matieland, South Africa}
\date{April 2009}
\begin{abstract} 
When smooth, zero-on-average, periodic magnetic and electric fields are applied to a 
carbon mono-layer (graphene), a gap between the valence and conduction band is introduced. 
Here this gapped state is studied analytically.
It is found that it does not correspond to a band insulator: 
a constant electric field induces a quantized Hall current 
even though the magnetic flux through the sample is zero and there are no Landau levels.
The phenomenon is of the same type as that 
discovered by Haldane for a graphene sample in a periodic magnetic field that is not smooth, 
i.e. varies rapidly on the scale of the graphene lattice constant. 
The effect can be explained in terms of the
topological theory of Thouless, Kohmoto, Nightingale and den Nijs.
For the system studied
in this paper, an explanation in terms of simple physical principles is also presented.
Thus some of the mystery is taken out of the apparently strange phenomenon of a Hall 
effect without magnetic flux. Furthermore, Haldane's model requires control over
external magnetic fields on length scales less than an angstrom and is therefore
hard to realize experimentally. For the model studied here, control 
over external fields on length scales that are larger by two orders of magnitude or more 
is sufficient. The model is therefore more amenable to experimental realization.
\end{abstract}
\pacs{72.15.-v, 73.43.-f \hfill NITheP-09-08}
\maketitle

The filled conduction band and the empty valence band of a carbon mono-layer (graphene)
touch at two inequivalent points in the Brillouin zone.\cite{Cas09} Regions of the Brillouin zone
in the vicinity of these points are called valleys. 
In each of the two valleys, low energy excitations are
described by a two-dimensional massless Dirac equation. As a consequence, electrons in graphene
can propagate through electrostatic potential barriers. In the jargon of relativistic
quantum mechanics, this is known as Klein tunneling.\cite{Kat06,Been07} When a gap is induced between 
the valence and conduction band, Klein tunneling is suppressed and the electronic
properties of graphene change radically. Several distinct situations have
been uncovered: 

Recently a method for producing a band insulating state was demonstrated experimentally.\cite{Zhou07} 
It involves growing the graphene sample on an 
appropriate substrate that induces a potential energy difference between the
two triangular sublattices that constitute the sample's honeycomb lattice.
In a seminal paper by Haldane,\cite{Hal88} it was demonstrated that
graphene's honeycomb lattice also supports a fundamentally different kind of gapped state
which I will call anomalous. 
A signature of the anomalous state is a Hall effect 
without Landau levels and in a zero average magnetic field. 
The resulting Hall conductivity is a topological invariant.\cite{Hal04}

In Haldane's model, the anomalous state is realized
by the combined effect of next nearest neighbor hopping together with a magnetic field
that has the same periodicity as the honeycomb lattice and zero flux through a 
lattice unit cell. Because of the very short length scales ($\sim1$\AA) on which this
magnetic field must be controlled,
the experimental realization of Haldane's model does not seem feasible. 
Recently, Kane and Mele \cite{Kane05} pointed out that the spin orbit interaction can
generate an anomalous gapped state in graphene without the need for any external magnetic
field. In this case, since time-reversal symmetry is not broken, no Hall effect with
electric charge current is produced. Rather a spin-Hall effect occurs. However,
the effect may be hard to observe because of the smallness of
the spin orbit interaction in graphene.  

A gapped state can also occur when periodic magnetic and electric fields that are 
smooth on the scale of the graphene lattice constant are applied jointly to a graphene
sample. Since these fields have a periodicity $\gg 1$\AA, experimental
realization is more feasible than in the case of Haldane's model.   
In this paper I show analytically that the resulting gapped state is of the same anomalous type
as in Haldane's model. 
Apart from being experimentally relevant, the system studied here has another interesting
feature. Its Hall effect can be explained without recourse to the 
topological theory. Because of this it is possible in the present instance to identify a physical
mechanism responsible for the Hall effect. This hopefully takes some of the mystery out of
the general but abstract topological theory.\cite{Kar54,Thou82,Thou85,Sim83a,Sim83b,Koh84,Osh94,Ono01,Hal04}

A gapped graphene state that results from smoothly varying fields was first revealed
in the numerical study of Ref. \onlinecite{Gui08} (see Figure 2, last frame, of that work) 
and is also mentioned in Ref. \onlinecite{Weh08}. The effective magnetic field studied 
in these references is produced by lattice deformations. The single-valley physics
of such a field is identical to that of a real magnetic field. The analytical results
for a single valley that is presented here confirm and explain numerical results obtained in
Refs. \onlinecite{Gui08} and \onlinecite{Weh08}. 
A caveat is warranted though:
Despite the close similarity between a graphene system with a real magnetic field and 
one with a deformed lattice, there is a fundamental difference. The gapped state produced
by deformations is not anomalous. The reason is that for deformation induced fields,
the anomalous behavior of the two valleys cancel each other, while for real magnetic fields,
the behavior is the same in both valleys and there is no cancellation.  

Several other publications deal with systems similar to that studied here. In Refs. \onlinecite{Kra96}
and \onlinecite{Tai08} the two-dimensional
non-relativistic Schr\"odinger equation with a periodic magnetic field that is zero on average was
studied. It is interesting to compare this system to the graphene system in the
limit of strong magnetic fields, when the magnetic length is much smaller than the periodicity
of the fields. In this regime the anomalous Hall effect\cite{foot1} was found to disappear in the case
of the Schr\"odinger equation.\cite{Tai08} It remains present in graphene as is shown here.
Another relevant study\cite{Brey09} dealt
with graphene in a periodic electric field alone. It was found that a strong electric field
can induce additional touching points between graphene's valence and conduction bands.
In the present work, only the case of weak electric fields will be considered,
so that additional touching points do not arise. What happens when both the electric
and magnetic fields are strong, or when the electric field is strong but the magnetic field weak,
is a topic for further investigation. 

There are also at least two relevant works in the carbon nanotube 
literature. In Ref. \onlinecite{Lee03} the low energy spectrum of a carbon nanotube in a magnetic field
that is transverse to the axial direction is calculated analytically. 
In Ref. \onlinecite{Bel07} the same
system is considered and shown to support a quantized Hall current in the axial direction.
The nanotube system can be considered a special 
one-dimensional case of the general two dimensional system considered below. In order
to induce a gap in the nanotube spectrum, an electric field must be applied parallel
to the magnetic field. This was not considered in the cited references. Consequently, the Hall effect
found in Ref. \onlinecite{Bel07} is fundamentally different from the Hall effect considered here. 
This will be discussed in more detail below.
However some results that hold for the system discussed in this work should also
hold for the nanotube system. Where there should be agreement between results in this paper
and in the nanotube studies, agreement is found.

The plan of the paper is as follows: In Sec. \ref{one} the system to be studied is defined mathematically.
In Sec. \ref{two} its zero energy eigenstates are found when the electric field is
zero, in which case there is no gap. In Sec. \ref{near_zero} the low energy description of the gapped
state is developed. In Sec. \ref{four} the topological theory of the anomalous Hall effect is
briefly reviewed and then applied to the graphene system in order to calculate the quantized Hall 
conductivity. In Sec. \ref{five} the anomalous Hall effect is considered again, this time
without invoking the topological theory. In Sec. \ref{six} an example is presented in order
to illustrate the general results of the previous sections. Conclusions are presented in Sec. \ref{seven}.
  
\section{Statement of the problem}
\label{one}

The inter-atomic distance between carbon atoms in graphene is approximately $1.42$\AA.
I consider a graphene sample in the presence of static electromagnetic fields $U$ and $\bm A$
that vary smoothly on this length scale. A long-wavelength description is therefore appropriate.
This involves the Dirac Hamiltonian
\begin{equation}
H=v\bm{\sigma}\cdot\left[-i\hbar\bm{\partial}_{\bm{r}}+e\bm{A(\bm{r})}\right]+U(\bm{r}).
\end{equation}
where $\bm{\sigma}=(\sigma_x,\sigma_y)$ are standard Pauli matrices,
$\bm{r}=(x,y)$, and $\bm{\partial}_{\bm{r}}=(\partial_x,\partial_y)$.
Close to the charge neutrality point, electrons in graphene obey the Dirac equation
$\varepsilon\Psi=H\Psi$. In principle two species of Dirac fermion should be distinguished
because the low-energy spectrum of graphene consists of two sets of Dirac cones.
However, since the fields $\bm A$ and $U$ are smooth, no scattering between species is
possible. Furthermore, by employing here the valley-isotropic representation,\cite{Been07} the Hamiltonians
for the two valleys are rendered identical, not only in form, but also in the actual values assumed
by the parameters. An index that distinguishes between valleys is therefore omitted. 
Results obtained will apply equally to electrons in both valleys unless otherwise indicated.

I consider a magnetic field 
$B\hat{\bm{z}}=\bm{\partial}_{\bm{r}}\times\bm{A}$ and a scalar potential $U$
that are periodic in space with the same periodicity and that average to zero.
The periodicity of the fields define a Bravais lattice with 
a primitive unit cell denoted $UC$ and basis vectors $\bm{a}_k$, $k=1,\,2$, such that  
$B(\bm{r}+\bm{a}_k)=B(\bm{r})$ and similarly for $U$. 

In what follows, any function that has this property
will be called $UC$-periodic.
Furthermore, whenever the term `lattice' is used below, it will
refer to the magnetic field lattice, not the crystal lattice of the carbon atoms
in graphene. The two lattices are entirely distinct and only the former is relevant
for present purposes. 
As a consequence of the smoothness of $\bm A$,
$UC$ is much larger than the unit cell of the graphene crystal lattice.   
Because $B$ averages to zero, the magnetic flux through $UC$ is
zero. 

The graphene sample is assumed to be much larger than $UC$ and only bulk effects will be 
considered. It is therefore convenient to impose periodic boundary conditions
\begin{equation} 
\Psi(\bm{r})=\Psi(\bm{r}+\Omega\bm{a}_k),~~k=1,2,\label{eq_bc}
\end{equation}
where $\Omega\gg1$ is an integer. All functions with this property will be called
sample-periodic. Note that because $\Omega$ is an integer, any $UC$-periodic function
is also sample-periodic.  

A reciprocal lattice is associated with the real space lattice. The basis vectors
$\bm{b}_j$, $j=1,\,2$ of the reciprocal lattice are defined through the equations 
$\bm{b}_j\cdot\bm{a}_k=2\pi\delta_{j,k}$. 
The Brillouin zone is a primitive unit cell of the reciprocal lattice, and will be denoted $BZ$.
It is convenient to define the Fourier components
\begin{equation}
B_{mn}=l^{-2}\int_{UC} dr^2\,e^{-i\bm{k}_{mn}\cdot\bm{r}}B(\bm{r}),\label{eq_bmn}
\end{equation}
of the magnetic field. Here $\bm{k}_{mn}=m\bm{b}_1+n\bm{b}_2$ with $m$ and $n$ integers 
and $l^2=|\bm{a}_1\times\bm{a}_2|$ is 
the area of $UC$. Note that $B_{00}=0$ due to 
the zero-flux condition imposed on $B$ and
$B_{m,n}=B_{-m,-n}^*$ because $B(\bm{r})$ is real.

\section{Zero energy modes}
\label{two}
The task is now to analyze the low-energy (i.e. small~$|\varepsilon|$) spectrum of $H$.

For $U=0$, zero energy eigenstates $\Psi=(\phi_+,\phi_-)$ of $H$ solve
\begin{equation}
\left[\partial_x\pm i\partial_y+i(\alpha_x\pm i\alpha_y\right)]\phi_\pm=0,\label{eq_zero_eigs}
\end{equation}
with $\bm{\alpha}=e\bm{A}/\hbar$.  
Following Jackiw,\cite{Jac84} 
I work in the Coulomb gauge where $\bm{\partial}_{\bm{r}}\cdot\bm{A}=0$. 
Then
Eq.~(\ref{eq_zero_eigs}), subject to the sample-periodic boundary conditions of Eq.~(\ref{eq_bc}), 
is solved by 
\begin{equation}
\phi_\pm(\bm{r})=c_\pm\exp[\pm F(\bm{r})],\label{eq_phipm}
\end{equation} 
where $c_\pm$ are arbitrary constants. The real function $F$ is $UC$-periodic 
and satisfies
\begin{equation} 
\bm{\partial}_{\bm{r}}^2F=eB/\hbar.\label{eq_d2F}
\end{equation}
In terms of the Fourier components $B_{mn}$ of the magnetic field, $F$ is given by
\begin{equation}
F(\bm{r})=-\frac{e}{\hbar}\sum_{mn\not=0}\frac{B_{m,n}}{|\bm{k}_{mn}|^2}e^{i\bm{k}_{mn}\cdot\bm{r}}.
\label{eq_F}
\end{equation}
 
One demonstrates uniqueness of the solutions in Eq. (\ref{eq_phipm}) as follows.
Consider the equation for $\phi_-$. Suppose that $\tilde{\phi}_-(\bm{r})$ solves Eq.~(\ref{eq_zero_eigs})
and is sample-periodic as required by the boundary conditions.
Define a function $g(\bm{r})=\tilde{\phi}(\bm{r})e^{F(\bm{r})}$. Since $F$ is $UC$-periodic, $g$ is sample-periodic. 
The function $g$ satisfies the Cauchy-Riemann equation
$(\partial_x-i\partial_y)g=0$. 
Thus $g$ is an analytical function of the complex variable $z=x+iy$. As a consequence $g$ is either constant
or unbounded in some directions at large $\bm{r}$. Unboundedness is incompatible with $g$ being sample-periodic
and therefore $g$ must be constant so that $\tilde{\phi}_-(\bm{r})=ge^{-F(\bm{r})}$ is proportional to 
$\phi_-(\bm{r})=c_-e^{-F(\bm{r})}$. The uniqueness of $\phi_+$ is demonstrated in the same manner.

The Dirac equation $H\Psi=E\Psi$ with $U=0$ therefore has exactly two zero energy solutions which I will
take to be
\begin{equation}
\Psi_{+,0}(\bm{r})=\frac{N_+}{\Omega}\left(\begin{array}{c}e^{F(\bm{r})}\\0\end{array}\right),
\hspace{3mm}
\Psi_{-,0}(\bm{r})=\frac{N_-}{\Omega}\left(\begin{array}{c}0\\e^{-F(\bm{r})}\end{array}\right),
\label{eq_sols2}  
\end{equation}
with $N_\pm=\left[\int_{UC}dr^2\,\exp\pm2F(\bm{r})\right]^{-\tfrac{1}{2}}$ 
ensuring normalization of $\Psi_{\pm,0}$ to unity over the sample.
The two eigenstates have opposite sublattice polarization: In valley $K$, $\Psi_{+,0}$ is
$A$ sublattice polarized while $\Psi_{-,0}$ is $B$-sublattice polarized. Since the
valley isotropic representation is employed, the roles of the sublattices are reversed
in the $K'$ valley. Hence, the state $\Psi_{+,0}$ in the $K'$ valley is $B$ polarized while
the $K'$ state $\Psi_{-,0}$ is $A$ polarized. The result is consistent with the Atiyah Singer index
theorem. It states that in a given valley the difference between the number of zero energy eigenstates
that are $A$ polarized and $B$ polarized equals the total flux through the sample.\cite{Pac07}
Note also that because $F$ is $UC$-periodic, the same is true for $\Psi_{\pm,0}$, even though only sample-periodicity
is imposed on $\Psi_{\pm,0}$ by the boundary condition [Eq.~(\ref{eq_bc})]. 

The solutions in Eq.~(\ref{eq_sols2}) were found without invoking Bloch's
theorem. I now briefly re-examine them from this point of view. Since the Hamiltonian
$H$ is periodic, an eigenbasis exists in which all eigenfunctions of $H$ are of the form
\begin{equation}
\Psi_{n,\bm{k}}(\bm{r})=\frac{1}{\Omega}e^{i\bm{k}\cdot\bm{r}}\psi_{n,\bm{k}}(\bm{r}),
\end{equation}
with $\bm{k}\in BZ$ and $n$ a discrete index. The functions $\psi_{n,\bm{k}}(\bm{r})$ are $UC$-periodic
and will be referred to as Bloch states. 
They satisfy $H(\bm{k})\psi_{n,\bm{k}}=\varepsilon_{n,\bm{k}}\psi_{n,\bm{k}}$ where $\varepsilon_{n,\bm{k}}$ 
are the energies of the eigenstates. The effective Hamiltonian $H(\bm{k})$ is given by   
\begin{equation}
H(\bm{k})=v\bm{\sigma}\cdot\left(-i\hbar\bm{\partial}_{\bm{r}}+\hbar\bm{k}+e\bm{A}\right)+U=H+\hbar v\bm{\sigma}\cdot\bm{k}.
\end{equation}
Normalizing $\psi_{n,\bm{k}}(\bm{r})$ to unity over $UC$ ensures that $\Psi_{n,\bm{k}}$ is normalized to unity over the whole sample.
The zero-eigenstates of Eq.~(\ref{eq_sols2}) are $UC$-periodic, and therefore correspond to $\bm{k}=0$ solutions
in this labeling scheme. The corresponding normalized zero-energy eigenstates of $H(\bm{k}=0)$ (with $U=0$) are
$\psi_{\pm,0}=\Omega\Psi_{\pm,0}$.

\section{Near Zero Eigenstates and a non-zero but small scalar potential} 
\label{near_zero}
The near-zero energy eigenstates in the vicinity of $\bm{k}=0$ can be studied by treating
the term $\hbar v\bm{\sigma}\cdot\bm{k}$ in $H(\bm{k})$ perturbatively. The effect of a weak periodic scalar potential
$U$ can be treated perturbatively at the same time. 

To leading order in $\bm{k}$ and $U$, the problem is solved by Bloch states of the form
\begin{equation}
u_{\eta,\bm{k}}(\bm{r})=\chi_{\eta,\bm{k}}^+\psi_{+,0}(\bm{r})+\chi_{\eta,\bm{k}}^-\psi_{-,0}(\bm{r}),
\end{equation}
where $\chi_{\eta,\bm{k}}=(\chi_{\eta,\bm{k}}^+,\chi_{\eta,\bm{k}}^-)$ satisfies 
$h(\bm{k})\chi_{\eta,\bm{k}}=\varepsilon_{\eta,\bm{k}}\chi_{\eta,\bm{k}}$.
Here $\eta=\pm$ labels the bands ($-$ for the valence band and $+$ for the conduction band), 
$\varepsilon_{\eta,\bm{k}}$
is the energy of the state with label $\eta,\bm{k}$ and $h(\bm{k})$ is the Hamiltonian $H(\bm{k})$ projected
onto the zero eigenspace. It has the form of a $2\times2$ Dirac Hamiltonian in k-space:  
\begin{align}
&h(\bm{k})_{\eta\eta'}=\int_{UC}dr^2\,\psi_{\eta,0}(\bm{r})^\dagger
\left[\hbar v\bm{\sigma}\cdot\bm{k}+U(\bm{r})\right]\psi_{\eta',0}(\bm{r}),\nonumber\\
&\implies h(\bm{k})=\hbar \tilde{v}\,\bm{\sigma}\cdot\bm{k}+\mu\,\sigma_z
\end{align}
where a term proportional to identity is omitted because it simply leads to
a redefinition of the zero-energy.
Here $\tilde{v}=l^2N_+N_- v$ is the renormalized Fermi velocity, 
and $l^2$ is the area of $UC$.
The effective mass $\mu$ is given by
\begin{equation}
\mu=\tfrac{1}{2}\int_{UC}dr^2~ U(\bm{r})\left(N_+^2e^{2F(\bm{r})}- N_-^2e^{-2F(\bm{r})}\right),\label{eq_mass}
\end{equation}
The signs of $U(\bm{r})$ and $F(\bm{r})$ are  
said to be correlated when there is a large overlap between regions of positive $U$ and positive $F$
and anti-correlated when there is a large overlap between regions of positive $U$ and negative $F$.
When the signs of $U$ and $F$ are either correlated or anti-correlated,
the two terms in the integrand tend not to cancel and $\mu$ is non-zero. 

The leading order perturbative expressions for the Bloch state $u_{\pm,\bm{k}}(\bm{r})$ 
in the vicinity of $\bm{k}=0$ is 
\begin{equation}
u_{\pm,\bm{k}}(\bm{r})=\tfrac{1}{\sqrt{2}}\left[\sqrt{1\pm\tfrac{\mu}{|\varepsilon_{\pm,\bm{k}}|}}\psi_+(\bm{r})
+\sqrt{1\mp\tfrac{\mu}{|\varepsilon_{\pm,\bm{k}|}}}e^{i\theta}\psi_-(\bm{r})\right],
\label{eq_u}
\end{equation}
with $\theta=\arg(k_x+ik_y)$ and $\varepsilon_{\pm,\bm{k}}=\pm\sqrt{(\hbar v|\bm{k}|)^2+\mu^2}$ the energy
of the state.
  
It is worth emphasizing here that the same effective Hamiltonian $h(\bm{k})$, and therefore
the same mass term is induced in both valleys. This is to be contrasted with the
mass term that arises when a staggered on-site potential is applied to otherwise clean graphene,
by for instance placing the sample on an appropriate substrate.\cite{Zhou07}
If the staggered potential takes on the value $m$ on the $A$ sublattice and $-m$ on the $B$ sublattice,
then a mass term $m\sigma_z$ is induced in valley $K$ while a mass term $-m\sigma_z$ is induced
in  valley $K'$.\cite{foot2}   

\section{The anomalous Hall effect}
\label{four}

In the previous section I demonstrated that there is a gap in the spectrum and that a low energy
description in terms massive Dirac fermions is possible. 
The gap has several important consequences. Firstly, low energy electrons can be localized in selected
regions of a graphene sample by inducing a gap in regions where electrons are to be excluded. The gap
can be controlled by electromagnetic fields that are smooth on the scale of the graphene lattice constant,
and there is no need to break the sub-lattice inversion symmetry.\cite{Zhou07}
Secondly, since one has control over the sign of the mass, it is possible to produce a sample in which 
the sign of the mass differs in different spatial regions. At the interface between two such regions,
chiral edge states are expected to appear.\cite{Kane05}
Thirdly, as is the case in Ref. \onlinecite{Hal88}, an electric field $\bm{E}=E\hat{y}$ applied to the
bulk of the sample (the $y$-direction is chosen arbitrarily) will produce a quantized current $j_x=\sigma_{xy}E$
in the direction perpendicular to it. This is remarkable since the magnetic field through the sample averages
to zero. 

Here I briefly review the general theory for this phenomenon, which is known as the anomalous Hall 
effect.\cite{Kar54,Thou82,Thou85,Sim83a,Sim83b,Koh84,Osh94,Ono01} 
It states that the Hall conductivity $\sigma_{xy}$ may be non-zero, even in the absence
of a net magnetic flux through the sample, provided time-reversal invariance is still broken. Furthermore
it is quantized in units of $e^2/h$ and can be written in terms of a set of integer
topological invariants as $\sigma_{xy}=(e^2/h)\sum_n {\rm Ch}_n$.\cite{Thou82,Thou85,Sim83a,Sim83b,Koh84,Osh94,Ono01}
Here $n$ is the discrete index
that labels the energy bands and the sum ranges over all occupied bands. (The theory assumes
that the Fermi energy is in a gap between bands so that there are no partially filled bands.) 
The invariant ${\rm Ch}_n$ is known as a Chern number, and
can be calculated if the Bloch state for band $n$ is known
throughout the whole Brillouin zone. Since 
only the Bloch states in the vicinity of $\bm{k}=0$ and only for the two bands that touch at
$\varepsilon=0$ were found in Sec. \ref{near_zero}, direct calculation of the ${\rm Ch}_n$ is not feasible here.

Fortunately, there is a way to work around this problem. 
The Chern numbers can only change when bands touch.\cite{Sim83a} 
Before and after a touching of bands, they differ by an integer.
Only the Chern numbers of bands involved in the touching change.
Furthermore, there is a sum-rule relating the Chern numbers before and after a band-touching.
Suppose that, by varying a parameter in the Hamiltonian, 
one induces bands $n$ and $n+1$ to touch before moving apart again.    
Let ${\rm Ch}_m^<$ (${\rm Ch}_m^>$)  be the Chern numbers before (after) the bands touched. 
Then ${\rm Ch}_n^<+{\rm Ch}_{n+1}^<={\rm Ch}_n^>+{\rm Ch}_{n+1}^>$, i.e. the sum of the two Chern numbers is preserved.
If bands $n$ and $n+1$ are both occupied, $\sigma_{xy}$ stays constant although
${\rm Ch}_n$ and ${\rm Ch}_{n+1}$ change. Thus $\sigma_{xy}$
only changes if the Fermi energy lies in the gap between bands $n$ and $n+1$.

In order to calculate the change in Chern numbers, one only needs information about the
bands that touch and only in regions of the Brillouin zone that are close to the points where 
the bands touch.\cite{Osh94,Ono01} This information may be obtained by means of a perturbative
expansion in the wave vector $\bm{k}$ and the external parameter that controls the closing of
the gap. (The general procedure is the same as was used in Sec. \ref{near_zero} of this text.) If
the first order coefficients do not vanish, the touching point is characterized by
a Dirac fermion. Its effective $2\times2$ Hamiltonian is brought into a standard form
(essentially the valley-isotropic representation) by means of appropriate $2\times2$
unitary transforms. It turns out that the change in Chern numbers is completely determined
by the sign of the mass term in this representation.   
If the bands touch at several points in $\bm{k}$ space simultaneously,
then each touching point gives an independent contribution.    

For the graphene system I am considering, this leads to a result 
\begin{equation}
{\rm Ch}_\pm^>-{\rm Ch}_\pm^<=\pm\{\rm{sign}[\mu^>]-\rm{sign}[\mu^<]\}.\label{eq_delta_chern}
\end{equation}
Here $+$ ($-$) refers to the conduction (valence) band and
$\mu^<$ ($\mu^>$) is the mass induced by the scalar potential before (after) the bands touched. 
There are two contributions to the result, one from each touching point. Since the same mass term appears
at both touching points (in the valley-isotropic representation), the contributions from the two touching points
are the same. If the sign of the mass were opposite in the two valleys, as is the case for
a substrate-induced mass, the two contributions would have canceled. 

There are two ways to
change the sign of the mass term [cf. Eq. (\ref{eq_mass})]. One may either change the sign of the external potential
$U$, or one may change the sign of the magnetic field and hence of $F$. According to the theory, 
both methods give the same change in the Hall conductivity,
$\Delta\sigma_{xy}=\Delta\sigma_{xy}(U\to-U)=\Delta\sigma_{xy}(B\to-B)$. 
The transformation $B\to-B$ is equivalent to time-reversal,
which changes the sign of the Hall conductivity: $\sigma_{xy}=-\sigma_{xy}$. 
Therefore $\Delta\sigma_{xy}=2\sigma_{xy}$. Thus, knowing the change in $\sigma_{xy}$ when the mass changes sign,
one also knows the actual value of $\sigma_{xy}$. 
 
If $\sigma_{xy}^\pm$ is the Hall conductivity of a 
sample in which the mass term has sign $\pm$ then $\sigma_{xy}^+-\sigma_{xy}^-=-2e^2/h$. Therefore,
according to the above argument
\begin{equation}
\sigma_{xy}^{\pm}=\mp \frac{e^2}{h}.\label{eq_hall}
\end{equation}

\section{Origin of non-zero Hall conductivity}
\label{five}

Suppose that the topological theory discussed in the previous section was not known. 
Would it have been possible to show that the graphene system studied here displays a
Hall effect, without at the same time essentially deriving the topological theory?           
As I show in this section, the answer is `yes'. In the process, the underlying physical
mechanism that here produces the anomalous Hall effect is identified. 

The starting point of the argument is to consider the two zero-energy solutions 
$\Psi_{\pm,0}\propto e^{\pm F}$ that are obtained when $U=0$. A key feature of these wave functions
is that the sign of $F$ is anti-correlated with that of $B$ i.e. $F$ is negative on average where the
magnetic field points in the positive $z$ direction and positive on average where the magnetic field points
in the negative $z$ direction. To prove this assertion I show that the number
$P=(e/\hbar)\int_{UC}dr^2\,B(\bm{r})F(\bm{r})$
is negative. The proof is elementary. From Eq.~(\ref{eq_d2F}) follows that 
$P=\int_{UC}dr^2\,B(\bm{r})\bm{\partial}_{\bm{r}}^2B(\bm{r})$.
Using integration by parts, and invoking Gauss's theorem I find
\begin{equation}
P=\underbrace{\oint_{\partial UC}d\bm{r}_\perp\cdot B\bm{\partial}_{\bm{r}}B}_{=0}
-\int_{UC}dr^2\,|\bm{\partial}_{\bm{r}}B|^2<0.\label{eq_p_negative}
\end{equation}
The integration in the first term is around the border of $UC$ and $d\bm{r}_\perp$ is normal
to the border (in the $x-y$ plane) pointing outwards and yields zero because
$B$ is $UC$-periodic. The result [Eq. (\ref{eq_p_negative})] implies that a particle in the 
state $\Psi_{-,0}$ ($\Psi_{+,0}$)
is largely confined to regions where $\bm{B}$ points in the positive (negative) 
$z$ direction. In Sec. \ref{two} it was noted that the eigenstates $\Psi_{\pm,0}$,
have opposite sublattice polarization. The correlation between $B$ and $F$ then
implies that sublattice polarization is correlated with the sign of the magnetic field.
For instance, in valley $K$, regions where the magnetic field points in the negative
$z$ direction are $A$ sublattice polarized while regions where the magnetic field 
points in the positive $z$ direction are $B$ sublattice polarized. 
This is consistent with the numerical results presented in Figure 3 of Ref. \onlinecite{Weh08}.
There it was pointed out that the state $\Psi_{-,0}$ closely resembles 
a state in the zero Landau level of a graphene sample with a constant magnetic field
that points in the positive $z$ direction. This would have explained the existence of
a Hall effect, were it not for the state $\Psi_{+,0}$ that also belongs to the zero 
eigenspace. This state is similar to a zero Landau level state in a magnetic field pointing
in the negative $z$ direction, and it is responsible for a Hall effect that cancels
that of $\Psi_{-,0}$.

In order to have a Hall effect a perturbation that lifts the degeneracy between $\Psi_{+,0}$ and
$\Psi_{-,0}$ is needed. This is the role of the electrostatic potential $U$.
If the sign of $U$ is 
anti-correlated with that of $B$ it will raise the energy of the state $\Psi_{+,0}$
while lowering the energy of the state $\Psi_{-,0}$. This leads to
a gap with associated positive mass. A scalar potential $U$
whose spatial profile is correlated with that of $B$ will do the
opposite and leads to gap associated with a negative mass.

Turning to the expression [Eq.~(\ref{eq_u}] for the $U\not=0$ valence band Bloch states 
$u_{-,\bm{k}}$, one sees that the dominant term in the linear combination is proportional
to $e^{-{\rm sign}(\mu) F}$. As a result, valence band
states are confined to regions of positive $B$ when $U$ is anti-correlated with $B$, and
confined to regions of negative $B$ when $U$ is correlated with $B$. This is consistent
with the response of $\Psi_{\pm,0}$ to $U$ discussed above.
 
One thus arrives at the fundamental result that there is a finite positive (negative) 
magnetic flux through the region of space occupied valence 
band electrons when
the mass is positive (negative). The valence band will then experience a `usual' Hall
effect. An electric field $\bm{E}=E\hat{\bm{y}}$ will, through the Lorentz force, 
lead to a current in the direction $-{\rm sign}(\mu)\, \hat{\bm{x}}$. 
Electrons in the conduction band would have seen an opposite magnetic field and produced
an opposite Hall effect. However, the Fermi energy is in the gap between the two bands
so that no conduction band states are occupied. The sign obtained here
for the Hall current agrees with the result derived from the topological theory 
[cf. Eq. (\ref{eq_hall})]. 

So far I have said nothing about the quantization of the Hall current. Indeed, the
above argument does not prove that the anomalous Hall conductivity precisely equals $\mp e^2/h$.
It only indicates that $\sigma_{xy}$ is finite and predicts its sign. However, the
argument does relate the anomalous Hall effect in graphene with periodic electromagnetic
fields to the usual integer Hall effect in a constant magnetic field. It thus shows
that the quantization of $\sigma_{xy}$ has the same origin in both cases. 

Finally, a word on the nanotube system: The Hall effect found there\cite{Bel07} is
not produced by states localized where the magnetic field has a well-defined direction, 
although such states are of course also present in that system. 
Rather it is produced by higher energy states that are localized where the magnetic
changes sign. Thus, a different mechanism is responsible for the effect reported in 
Ref. \onlinecite{Bel07}. Other features also distinguish it from the Hall effect 
considered in the present work: It is not accompanied by a gap between the conduction
and valence bands. It is associated with two sets of states on opposite sides of the
circumference of the nanotube that carry current along the length of the nanotube but 
in opposite directions. The quantization is approximate.
This is to be contrasted with the following properties
of Hall effect discussed in the present work. It appears when the Fermi energy
is in the gap between the conduction and valence bands. All the current travels
in one direction throughout the whole sample. The Hall quantization of the
Hall conductivity is sharp.    

The conclusion is that the Hall effect discussed in this paper has not yet
been found in the context of carbon nanotubes. But is it there? 
The topological theory cannot answer this question. It only works
for systems that are much larger than the periodicity of the magnetic field, whereas the
length of the nanotube system in the circumferential direction is the same as the 
periodicity of the magnetic field. It is not clear if the intuitive argument 
presented in this section can be applied either. It also involves viewing the system with
a spatial resolution that is larger than the periodicity of the magnetic field.
The question therefore remains open. 
 
\section{Examples}
\label{six}
{\em Example 1:} As a simple example that illustrates the general results derived in this text, 
consider the magnetic field
$B(\bm{r})=B_0\left(\sin2\pi x/\lambda+sin 2\pi y/\lambda\right)$
with periodicity $\lambda$ in both the $x$ and $y$ directions.

It is convenient to define the magnetic length $l_m=\sqrt{\hbar/e|B_0|}$ and a
dimensionless constant $\beta=e\lambda^2|B_0|/\hbar=(\lambda/l_m)^2$.  
According to Eq.~(\ref{eq_F}), $F$ is given by
\begin{equation}
F(\bm{r})=-\frac{\beta}{4\pi^2}\frac{B(\bm{r})}{|B_0|}.
\end{equation}
This gives the clearest possible illustration
that $\Psi_{\pm,0}$ are localized to regions of positive or negative $B$. 
The degree of localization is controlled by the ratio $\lambda/l_m$.
 
\begin{figure}[tbh]
\begin{center}
\includegraphics[width=.8 \columnwidth]{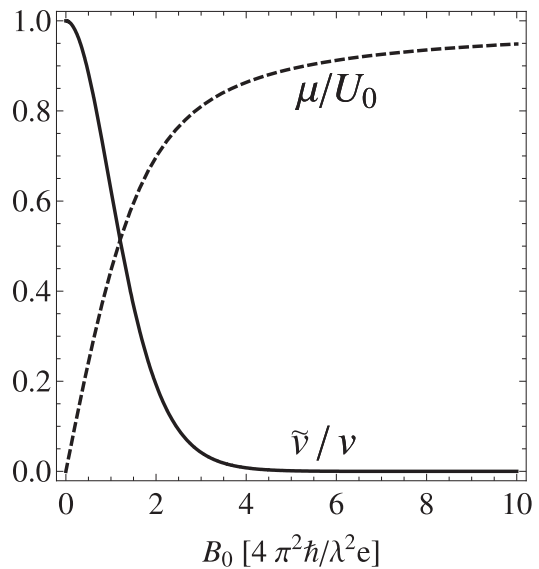}
\caption{The renormalized Fermi velocity $\tilde{v}$ (solid line) according to Eq. (\ref{eq_tildev})
and the mass $\mu$ (dashed line) according to Eq. (\ref{eq_muii})
for the model with magnetic field 
$B(\bm{r})=B_0\left(\sin2\pi x/\lambda+sin 2\pi y/\lambda\right)$ perpendicular
to the graphene plane. 
\label{fig0}}
\end{center}
\end{figure}

The renormalized Fermi velocity can be calculated explicitly as
\begin{equation}
\tilde{v}=\frac{v}{I_0(\beta/4\pi^2)^2},\label{eq_tildev}
\end{equation}
where $I_0$ is the zero'th modified Bessel function of the 
first kind. When $\beta\gg1$ or equivalently
$l_m\ll\lambda$, the asymptotic form $\tilde{v}\simeq \beta e^{-\beta/2\pi^2}v/2\pi$ is valid.
In this limit the Fermi velocity becomes very small and bands become nearly dispersionless 
in the region of $\bm k=0$. The result is very similar to the nanotube result.\cite{Lee03}
This is due to the fact that the spatial profile
of the magnetic field in the nanotube case is also sinusoidal.
In Figure \ref{fig0} the solid line shows the
renormalized Fermi velocity (solid line) plotted as a function of the magnetic field 
strength $B_0$. 

As was pointed out in Ref. \onlinecite{Gui08} and \onlinecite{Weh08}
the flat bands are reminiscent of Landau levels. 
It should however be pointed out that flat bands do not guarantee a
Hall effect: The non-relativistic Schr\"odinger 
equation in the presence of the magnetic field of this example was studied numerically
in Ref. \onlinecite{Tai08}. Flat bands were found in the limit $\lambda/l_m\gg1$. For that system
the Chern numbers of flat bands were found to be zero, leading to a zero Hall conductivity.

Neither are the flat bands a necessary condition for the Hall effect to occur. The effect
also occurs for $l_m >\lambda$ when the renormalized Fermi velocity is of the same 
order of magnitude as the unrenormalized Fermi velocity. Counter-intuitively, it is 
not the flat bands, but the Dirac cones at the touching points between bands,
that lead to the Hall effect. 

Returning to the analysis of the example I now introduce a
scalar potential $U(\bm{r})=-U_0 B(\bm{r})/B_0$ with the same spatial profile
as the magnetic field. The induced mass is
\begin{equation}
\mu={\rm sign}(B_0)U_0\frac{I_1(\beta/4\pi^2)}{I_0(\beta/4\pi^2)}\simeq{\rm sign}(B_0)U_0,\label{eq_muii}
\end{equation} 
where $I_1$ is the first modified Bessel function of the 
first kind. The asymptotic form is valid in the same limit as before, 
namely when $l_m\ll\lambda$. This result is straight forward to interpret. In the
strong magnetic field limit the eigenstates become
exponentially well-confined to the maxima and minima of the magnetic field. These
regions are also the maxima and minima of $U$, so that the effective potential
seen by the eigenstates has magnitude $|U_0|$. The behavior of the mass $\mu$ as a function of the 
magnetic field strength $B_0$ is shown (dashed line) in Figure \ref{fig0}.

{\em Example 2:} Finally, I consider an example in which the magnetic and electric fields are constant
in one direction and periodic in the other. The full spectrum can be obtained analytically. 
The approximate formulas for the low energy spectrum in the previous sections can therefore
be tested against exact results.
 
The magnetic field and electrostatic potential are taken as
\begin{eqnarray}
\bm{B}&=&\hat{\bm{z}}\,B_0\lambda\sum_{n\in Z}\left[\delta(y-n\lambda)-\delta(y-(n+\tfrac{1}{2})\lambda)\right],\nonumber\\  
U&=&-U_0\lambda\sum_{n\in Z}\left[\delta(y-n\lambda)-\delta(y-(n+\tfrac{1}{2})\lambda)\right].\label{eq_1dbu}
\end{eqnarray}
The function $F$ is given by
\begin{equation}
F=\frac{\lambda^2}{2l_m^2}
\left(
\frac{1}{4}-\left|\frac{y_{\,\rm mod \,\lambda}}{\lambda}-\frac{1}{2}\right|
\right).
\end{equation}
According to the results presented in Sec. \ref{near_zero}, the renormalized Fermi velocity and mass are 
\begin{equation}
\tilde{v}
=\frac{\beta}{4\sinh(\beta/4)}v,~~\mu={\rm sign}(B_0)\frac{\beta U_0}{2},\label{eqmuv}
\end{equation}
with $\beta$ defined as in Example 1. Again the renormalized Fermi velocity is an exponentially 
decreasing function of $\beta$ and hence of $|B_0|$.

\begin{figure}[tbh]
\begin{center}
\includegraphics[width=.9 \columnwidth]{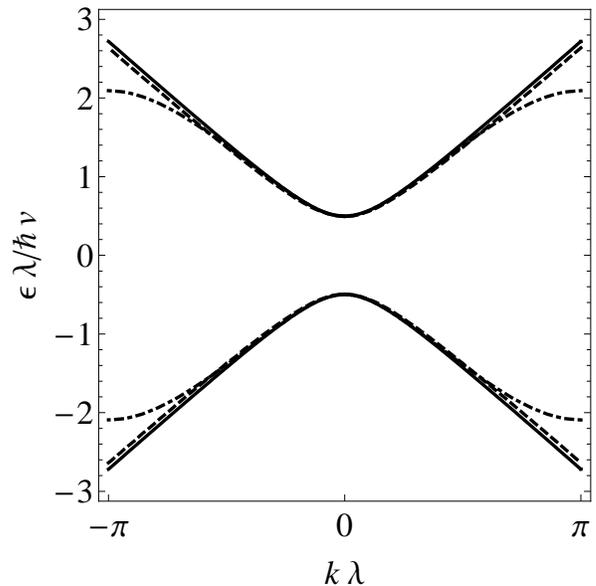}
\caption{
The low energy dispersion of the Dirac equation with the magnetic field and
electrostatic potential of Eq. (\ref{eq_1dbu}). The dashed line shows the exact result 
for the wave vector $\bm{k}$ chosen along the $x$ direction. The dot-dashed line shows the
exact result for the wave vector $\bm{k}$ along the $y$ direction. The solid line shows
the Dirac dispersion $\varepsilon=\pm\sqrt{(\hbar \tilde{v} |\bm k|)^2+\mu^2}$ with
$\mu$ and $\tilde{v}$ as in Eq. (\ref{eqmuv}). A magnetic length $l_m=\lambda/2$ 
and a potential strength $U_0=\hbar v/4\lambda$ was used.  
\label{fig1}}
\end{center}
\end{figure}

The full spectrum of the model can be obtained from the transcendental equation
\begin{align}
&\cos(k_y\lambda)=\cos(\tfrac{\lambda p_+}{2})\cos(\tfrac{\lambda p_-}{2})\nonumber\\
&-\frac{\sin(\frac{\lambda p_+}{2})}{p_+}\frac{\sin(\frac{\lambda p_-}{2})}{p_-}\left[\left(\frac{\varepsilon}{\hbar v}\right)^2-k_+k_
-\cos\left(\tfrac{2U_0\lambda}{\hbar v}\right)\right],
\end{align}
where $p_\pm=\sqrt{(\varepsilon/\hbar v)^2-k_\pm^2}$, $k_\pm=k_x\pm\lambda/2 l_m^2$, $\varepsilon$ is the energy and
the wave vector is $\bm{k}=k_x\hat{\bm{x}}+k_y\hat{\bm{y}}$. In Figure \ref{fig1} the exact dispersion of the two bands
closest to $\varepsilon=0$ is plotted in the $x$ and $y$ directions. On the same plot is also shown the hyperbolic dispersion
relation $\varepsilon=\pm\sqrt{(\hbar \tilde{v} |\bm k|)^2+\mu^2}$, with $mu$ and $\tilde{v}$ as calculated in Eq. (\ref{eqmuv}).
It is seen that the low energy dispersion relation is reproduced very well by the approximate formulas.

\section{Conclusion}
\label{seven} 
In this paper I showed analytically that
when a graphene sample is exposed to a periodic magnetic field that (a) is smooth on the scale of
the graphene lattice, and (b) is zero on average, the valence and conduction bands still
touch at two Dirac points, but the Fermi velocity is reduced. The reduction the same in all directions 
and is exponential in
the magnetic field strength. For simple models that introduce no other length scales beyond 
the magnetic length $l_m$ and the periodicity $\lambda$ of the magnetic field, the renormalized
Fermi velocity $\tilde{v}$ behaves as $\tilde{v}/v\propto\exp(c \lambda^2/l_m^2)$ with $c$
of order unity. The zero eigenspace is two-fold degenerate. A basis exists in which one
eigenstate confines particles to regions where magnetic field points in the positive $z$ direction
and to sublattice $A$ ($B$) in valley $K$, ($K'$). The other eigenstate does the opposite, 
i.e. confines particles to regions where the magnetic field points in the negative $z$ direction
and to sublattice $B$ ($A$) in valley $K$, ($K'$). An electrostatic potential that is either correlated
or anti-correlated with the magnetic field induces a gap between the valence and conduction bands.
These analytical results are consistent with previously obtained numerical results.\cite{Gui08,Weh08}
I showed that the gapped state supports a quantized Hall effect with $\sigma_{xy}=\pm e^2/h$. The positive sign
refers to the correlated case and the $-$ sign refers to the anti-correlated case. This is an instance
of the known phenomenon of a Hall effect in zero overall flux.\cite{Hal88} It can be explained in terms
of the topological properties of Bloch states of the occupied bands.\cite{Kar54,Thou82,Thou85,Sim83a,Sim83b,Koh84,Osh94,Ono01} 
In the present system I showed that 
the effect is also simply related to the usual integer Hall effect in a non-zero average magnetic field: 
In the presence of the electrostatic potential, the spatial region in which the filled valence band states
are localized, is permeated by a finite total magnetic flux. As a result, the valence band
electrons see a magnetic field with a non-zero average even though the magnetic field averages
to zero over the sample as a whole. 

\acknowledgments
This research was supported by the National Research Foundation (NRF) of South Africa.


\begin{thebibliography}{99}
\bibitem{Cas09} A. H. Castro Neto, F. Guinea, N. M. R. Peres, K. S. Novoselov, and A. K. Geim,
Rev. Mod. Phys. {\bf 81}, 109 (2009).
\bibitem{Kat06} M. I. Katsnelson, K. S. Novoselov, and A. K. Geim, Nature Phys. {\bf 2}, 620 (2006). 
\bibitem{Been07} C. W. J. Beenakker, Rev. Mod. Phys. {\bf 80}, 1337 (2008).
 
\bibitem{Zhou07} S. Y. Zhou, G.-H. Gweon, A. V. Fedorov, P. N. First, W. A. de Heer,
D.-H. Lee, F. Guinea, A. H. Castro Neto, and A. Lanzara, Nature Mat. {\bf 6}, 770, (2007).

\bibitem{Hal88} F. D. M. Haldane, Phys. Rev. Lett. {\bf 61}, 2015 (1988).

\bibitem{Hal04} F. D. M. Haldane, Phys. Rev. Lett. {\bf 93}, 206602 (2004).

\bibitem{Kane05} C. I. Kane and E. J. Mele, Phys. Rev. Lett. {\bf 95}, 226801 (2005).

\bibitem{Kar54} R. Karplus and J. M. Luttinger, Phys. Rev. {\bf 95}, 1154 (1954).

\bibitem{Thou82} D. J. Thouless, M. Kohmoto, M. P. Nightingale, and M. den Nijs, Phys. Rev. Lett. 
{\bf 49}, 405 (1982)

\bibitem{Thou85} Q. Niu, D. J. Thouless, and Y. Wu, Phys. Rev. B {\bf 31}, 3372 (1985).

\bibitem{Sim83a} J. E. Avron, R. Seiler, and B. Simon, Phys. Rev. Lett. {\bf 51}, 51 (1983).   

\bibitem{Sim83b} B. Simon, Phys. Rev. Lett. {\bf 51}, 2167 (1983).

\bibitem{Koh84} M. Kohmoto, Ann. Phys. (N.Y.) {\bf 160}, 343 (1984).

\bibitem{Osh94} M. Oshikawa, Phys. Rev. B {\bf 50}, 17357 (1994).

\bibitem{Ono01} M. Onoda and N. Nagaosa, J. Phys. Soc. Jpn. {\bf 71}, 19 (2002).

\bibitem{Gui08} F. Guinea, M. I. Katsnelson and M. A. H. Vozmediano, Phys. Rev. B {\bf 77}, 075422 (2008).

\bibitem{Weh08} T. O. Wehling, A. V. Balatsky, A. M. Tsvelik, M. I. Katsnelson, and A. I. Lichtenstein,
 Euro. Phys. Lett. {\bf 84}, 17003 (2008).
 
\bibitem{Kra96} A. Krakovsky, Phys. Rev. B {\bf 53}, 8469 (1996).

\bibitem{Tai08} M. Taillefumier, V. K. Dugaev, B. Canals, C. Lacroix, and P. Bruno, Phys. Rev. B {\bf 78}, 155330 (2008).

\bibitem{foot1} Unfortunately `anomalous Hall effect' is also the standard terminology for
another phenomenon in graphene, namely the unusual quantization of Landau levels. This is not
what the term refers to in this text. Rather it refers to a Hall effect in a zero average magnetic field.

\bibitem{Brey09} L. Brey and H. A. Fertig, arXiv:0904.0540 (2009).

\bibitem{Lee03} H.-W. Lee and D. S. Novikov, Phys. Rev. B {\bf 68}, 155402 (2003).

\bibitem{Bel07} S. Bellucci, J. Gonz\'alez, F. Guinea, P. Onorato, and E. Perfetto, 
J. Phys.: Condens. Matter {\bf 19}, 395017 (2007).
 
\bibitem{Jac84} R. Jackiw, Phys. Rev. Lett. {\bf 29}, 2375 (1984).

\bibitem{Pac07} J. K. Pachos and M. Stone, Int. J. Mod. Phys. B {\bf 21}, 5113, (2007).

\bibitem{foot2}
The sign of the mass term in a given valley depends on the representation.
All statements made in this text hold for the valley-isotropic representation. Care should be taken
when comparing with results in the literature since several other representations are also common. See
footnote 2 of Ref. \onlinecite{Been07}.

\end{thebibliography}
\end{document}